\documentclass[journal]{IEEEtran} 

\ifCLASSINFOpdf
\else
\fi

\usepackage{color}
\usepackage{amssymb}
\usepackage{amsthm}
\usepackage[cmex10]{amsmath}
\DeclareMathOperator{\E}{\mathbb{E}}

\DeclareMathOperator{\C}{\mathbb{C}}
\DeclareMathOperator{\R}{\mathbb{R}}

\newcommand {\Define} {\stackrel {\Delta} {=}  }

\newcommand{\mya}{\mathrel{\overset{\makebox[0pt]{{\tiny(a)}}}{=}}}
\newcommand{\myageq}{\mathrel{\overset{\makebox[0pt]{{\tiny(a)}}}{\geq}}}

\newcommand{\myb}{\mathrel{\overset{\makebox[0pt]{{\tiny(b)}}}{=}}}

\newcommand{\myc}{\mathrel{\overset{\makebox[0pt]{{\tiny(c)}}}{=}}}

\newcommand{\myd}{\mathrel{\overset{\makebox[0pt]{{\tiny(d)}}}{=}}}

\newcommand{\mye}{\mathrel{\overset{\makebox[0pt]{{\tiny(e)}}}{=}}}

\usepackage{bm}
\usepackage{mathtools}
\usepackage{fixltx2e}
\usepackage{epsfig,makeidx,color}
\usepackage{graphicx,dblfloatfix}
\usepackage{amsbsy}
\usepackage{amssymb}
\usepackage{euscript}
\usepackage{lipsum}
\usepackage{cite}
\usepackage{chngcntr}
\usepackage{lineno}
\usepackage[T1]{fontenc}
\usepackage{microtype}
\usepackage{wasysym}
\usepackage{footnote}
\newtheorem{theorem}{Theorem}

\newtheorem{proposition}{Proposition}

\newtheorem{remark}{\it Remark}

\usepackage[english]{babel}
    \usepackage[font=small,labelsep=space]{caption}
\usepackage{subcaption}    
\makeatletter
\def\citenoauxwrite#1{\begingroup
\@fileswfalse
\cite{#1}\relax
\endgroup}
\makeatother

\begin{document}

\title{Impact of Underlaid Multi-antenna D2D on Cellular Downlink in Massive MIMO Systems}
%
%
%

\author{Amit~Agarwal, 
~Sudarshan~Mukherjee and 
        ~Saif Khan~Mohammed 
\thanks{The authors are with the Department of Electrical Engineering, Indian Institute of Technology (I.I.T.) Delhi, India. Saif Khan Mohammed is also associated with Bharti School of Telecommunication Technology and Management (BSTTM), I.I.T. Delhi. Email: saifkmohammed@gmail.com. This work is supported by EMR funding from the Science and Engineering
Research Board (SERB), Department of Science and Technology (DST),
Government of India.}
}

\maketitle


\begin{abstract}
In this paper, we consider a massive multiple-input multiple-output (MIMO) downlink system underlaid with a network of multi-antenna D2D user equipments (UEs). Each D2D transmitter (Tx) uses all its antennas to beamform information towards its desired D2D receiver, which uses only a single antenna for reception. While beamforming at the D2D Tx reduces D2D interference to the neighbouring cellular UEs (CUEs), the cellular-to-D2D interference is also negligible due to highly directional beamforming at the massive MIMO base station. For the above proposed system, we analyze the average per-user spectral efficiency (SE) of CUEs ($R^{c,d}$) as a function of the D2D area spectral efficiency (ASE). Our analysis reveals that for a fixed D2D ASE ($R_0^{(d)}$) and fixed number of D2D antennas ($N$), with increasing density of D2D Txs ($\lambda$), $R^{c,d}$ increases (for sufficiently large $\lambda$) and approaches a fundamental limit $R_{\infty}^{c,d}$ as $\lambda \to \infty$. Also, $R_{\infty}^{c,d}$ depends on $R_{0}^{(d)}$ and $N$, only through the ratio $\frac{R_{0}^{(d)}}{N-1}$, i.e, for a given fundamental limit $R_{\infty}^{c,d}$, the D2D ASE can be approximately doubled with every doubling in $N$.
\end{abstract}


\vspace{-0.5 cm}

%


\section{Introduction}
%
%
%
%
The device-to-device (D2D) proximity services, initially conceived as a part of the 3\textsuperscript{rd} generation partnership project (3GPP), has attracted a lot of attention as one of the key $5$G technologies \cite{Lte,Lin2,Tehrani}. D2D communication enables peer-to-peer and location-based proximity services by providing direct communication link between mobile devices, instead of routing the traffic through the cellular base station (BS) \cite{Shalmashi,Lin2}. This feature not only reduces packet latency and improves energy efficiency, but also improves spectral efficiency by accommodating more number of user equipments (UEs) \cite{Andrews}. However introduction of D2D communication services underlaid in the existing cellular architecture is a challenging problem. Since underlaid D2D UEs share the spectral resources with the cellular UEs (CUEs), uncontrolled growth in the D2D network would severely degrade the cellular system performance. In recent years, massive multiple-input multiple-output (MIMO) systems have been favoured as a key $5$G cellular technology to counter the impact of multi-user interference (MUI) \cite{Marzetta2}. Introduction of D2D underlaid in massive MIMO cellular network therefore has been envisaged as a potential implementation scenario for D2D underlay in cellular systems.

 \par Most of the recent works in massive MIMO cellular systems with D2D underlay focus on the impact on the energy efficiency (EE) and spectral efficiency (SE) due to the presence of D2D UEs \cite{Lin,Shalmashi}. In \cite{Shalmashi}, single antenna D2D UEs underlaid in massive MIMO system have been studied for EE and average sum-rate (cellular rate plus D2D rate). In \cite{Lin} massive MIMO uplink with multi-antenna D2D underlay have been considered to improve D2D SE (through diversity combining at the D2D receiver) and reduce the D2D interference at the cellular BS for partial zero-forcing (PZF) receivers. From the results presented in these papers, however, it is difficult to explicitly infer the impact of underlaid D2D interference on the per-user cellular SE as a function of D2D information rate. Further in these papers, underlaid D2D communication has been considered only with massive MIMO uplink, based on the assumption that in the cellular downlink, the D2D-to-cellular interference would be very high for the neighboring cellular UEs \cite{Lin2,Lin}. \textit{This assumption however has not been investigated for underlaid multi-antenna D2D UEs, communicating in massive MIMO cellular downlink}. Therefore in this paper, we investigate the impact of underlaid multi-antenna D2D interference on the spectral efficiency of CUEs in massive MIMO downlink.

\par In this paper, we consider a massive MIMO cellular downlink, with underlaid multi-antenna D2D UEs. Each D2D UE is equipped with $N$ antennas. In massive MIMO downlink, the \textit{cellular-to-D2D interference is negligible} due to highly directional beamforming at the BS. We also propose a \textit{new} D2D communication strategy, where a D2D transmitter (Tx) uses all $N$ antennas to \textit{selectively} beamform information only towards its desired D2D receiver, which uses a \textit{single antenna} for reception. The resulting array gain allows us to reduce the D2D transmit power, thereby reducing the D2D interference power to the neighbouring CUEs. In this paper, for this above proposed D2D communication strategy, we characterize the impact of the D2D interference on the average per-user SE of CUEs as a function of the area spectral efficiency (ASE) of the underlaid D2D network. To the best of our knowledge, this is the first paper to report such a study.

\textit{Contributions}: The novel contributions presented in this paper are as follows: (i) for the above proposed system, using stochastic geometry tools, we derive closed-form expressions for the ASE of the D2D network and also for the average per-user SE of CUEs; (ii) for a fixed D2D ASE and fixed number of D2D Tx antennas ($N$), the average per-user SE of CUEs increases with increasing D2D pair density ($\lambda$), when $\lambda$ is sufficiently large and it approaches a fundamental limit as $\lambda \to \infty$; (iii) our analysis also reveals that for a fixed $N$ and given D2D ASE, $R_{0}^{(d)}$, this fundamental limit depends on ($N, R_{0}^{(d)}$) only through the ratio $\frac{R_{0}^{(d)}}{N-1}$. This suggests the interesting result that by doubling the number of D2D Tx antennas, $N$, the D2D ASE can also be doubled, while maintaining a fixed fundamental limit on the average per-user SE of CUEs. [\textbf{{Notations:}} $\C$ denotes the set of complex numbers. $\E$ denotes the expectation operator. $(.)^{\ast}$ is the complex conjugate operator. $\bm I_{N}$ represents the $N\times N$ identity matrix.]

\vspace{-0.5 cm}

\section{System Model}

\begin{figure}[t]
\vspace{-0.4 cm}
\includegraphics[width= 3 in, height= 1.8 in]{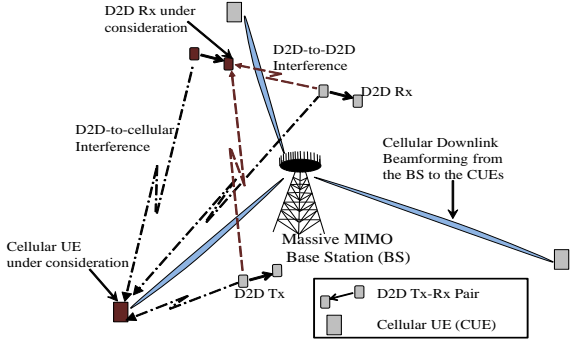}
\caption {Proposed Single Cell Massive MIMO Downlink with underlaid D2D network. Here the massive MIMO base station beamforms information signal towards the individual cellular UEs.}
\vspace{-0.5 cm}
\label{fig:d2dsysmodel}
\end{figure}

We consider a single cell time division duplexed (TDD) massive MIMO system, serving a fixed number of single-antenna cellular UEs (CUEs) in the same time-frequency resource. The cellular system is also underlaid with a network of multi-antenna D2D UEs.\footnote[1]{We assume that at any time slot, a UE can operate either as a CUE or as a D2D UE, but not as both.} The D2D UEs form D2D pairs, where each D2D UE can operate in one of the following two modes: (a) D2D transmitter (Tx) (all antennas are used for transmission); (b) D2D receiver (Rx) (single active antenna for reception).\footnote[2]{The Tx/Rx mode of operation can be assigned either through mutual cooperation of the UEs in a pair or by the BS.} Thus each D2D Tx-Rx pair forms a MISO (multiple-input single-output) system. In this paper, we assume that the UEs in a D2D pair are always at a fixed distance $D > d_0$ from each other ($d_0$ is the far-field distance from a transmitter) and $P_D$ is the total average power transmitted by a D2D Tx. In our system consideration, the D2D Txs actively transmit only during cellular downlink (see Fig.~\ref{fig:d2dsysmodel}). In this scenario, we study the impact of D2D interference on the average per-user cellular information rate, for a given area spectral efficiency (ASE) of the D2D network.

To determine the per-user spectral efficiency (SE) of CUEs we proceed as follows. Firstly we assume that all D2D pairs are uniformly distributed and hence the impact of D2D interference on any CUE is independent of the location of the CUE. To compute the D2D interference on a CUE, we model the locations of D2D Txs using a homogeneous Poisson Point Process (PPP) $\Phi \in \R^2$ of intensity $\lambda$, with the CUE under consideration at the origin. Since we are interested in the impact of D2D interference on the average cellular SE for a given D2D ASE, we also require an expression for the D2D ASE, which is a function of the D2D Tx density per unit area ($\lambda$), D2D Tx power ($P_D$) and the number of antennas in a D2D Tx ($N$). To compute the ASE, we \textit{consider} a D2D pair and evaluate the useful signal power received at the D2D Rx under consideration from the D2D Tx in that pair. Thus considering the D2D Rx under consideration to be at the origin, we model the corresponding D2D Tx at a fixed distance $D$ from it. To model the total D2D interference at the D2D Rx under consideration from all other D2D Txs, we model their locations randomly according to a homogeneous PPP $\Phi$ of intensity $\lambda$. We denote the D2D Tx-Rx pair under consideration as the $k = 0$-th D2D pair. The channel gain vector from the D2D Tx (of the $k$-th pair) to the D2D Rx (of the $k$-th pair) is given by $\bm g_k \Define (g_{k,1}, g_{k,2}, \cdots, g_{k,N})^T = \sqrt{\beta_k} \, \bm h_k \in \C^{N \times 1}$, where $g_{k,i}$ ($i = 1, 2, \ldots, N$) denotes the channel gain between the $i$-th antenna of the $k$-th D2D Tx and its corresponding D2D Rx. $\sqrt{\beta_k} > 0$ is the geometric attenuation (pathloss and shadow fading) and $\bm h_k \sim \mathcal{C}\mathcal{N}(0,\bm I_N)$ is the fast fading component having independent and identically distributed (i.i.d.) components.

\par We further assume that the $k$-th D2D Tx beamforms information symbol $u_k \sim\mathcal{C}\mathcal{N}(0,1)$ towards its corresponding D2D Rx by using maximum ratio transmission (MRT). Therefore the signal received at the D2D Rx under consideration (i.e. the $0$-th D2D Rx at origin) is given by\footnote[3]{Since we are operating in the cellular downlink, the cellular-to-D2D interference is negligible due to highly directional beamforming at the massive MIMO BS.}

\vspace{-0.6 cm}

\begin{IEEEeqnarray}{rCl}
\label{eq:rxsigd2d0}
y & = & \sqrt{P_D}||\bm g_0||u_0 + \underbrace{\sqrt{P_D}\sum\limits_{k \in \Phi} \frac{\bm g_k^{(d)T}\bm g_k^{\ast}}{||\bm g_k||} u_k +n}_{\text{Interference and Noise terms}},
\IEEEeqnarraynumspace
\end{IEEEeqnarray}

\vspace{-0.2 cm}

\noindent where $n \sim \mathcal{C}\mathcal{N}(0, \sigma^2)$ is the complex circular symmetric AWGN at the D2D Rx under consideration. Further $\bm g_k^{(d)} = \sqrt{\beta_k^{(d)}} \, \bm h_k^{(d)} \in \C^{N \times 1}$ is the channel gain vector from the $k$-th D2D Tx to the D2D Rx under consideration, where $\bm h_k^{(d)} \sim \mathcal{C}\mathcal{N}(0, \bm I_N)$ are i.i.d. and independent of $\bm h_k$. $\sqrt{\beta_k^{(d)}}>0$ is the pathloss and shadow fading component, which is modelled using the generalized pathloss model \cite{Goldsmith} as given below:

\vspace{-0.5 cm}

\begin{IEEEeqnarray}{rCl}
\label{eq:pathloss2}
\beta_k^{(d)} = \left({\lambda_c}/{4\pi d_0}\right)^2 \, ({1}/{l(r_k)}) \, e^{-\frac{\Xi_{\text{dB}}}{10}\ln{10}}.
\IEEEeqnarraynumspace
\end{IEEEeqnarray}

\vspace{-0.2 cm}

\noindent where $l(r_k) \Define [\max(1,r_k/d_0)]^{\alpha_d}$ \cite{Bai}. Here $\alpha_d >2$ is the pathloss exponent and $r_k$ denotes the distance between the $k$-th interfering D2D Tx and the $0$-th D2D Rx. $\lambda_c \Define c/f_c$, where $c = 3 \times 10^{8}$ m/s, $f_c$ is the carrier frequency and $\Xi_{\text{dB}}$ is a real zero mean Gaussian random variable with standard deviation $\sigma_{\text{db}}$ dB which models the lognormal shadowing. In \eqref{eq:rxsigd2d0}, the useful signal power for a given set of channel realizations is $\E[ \, |\sqrt{P_D} \, ||\bm g_0 \, || \, u_0|^2] = P_D \beta_0 ||\bm h_0||^2$. From \eqref{eq:rxsigd2d0} it also follows that the variance of the interference and noise terms is given by $P_D \sum_{k \in \Phi} \beta_k^{(d)} |\frac{\bm h_k^{(d) \, T} \, \bm h_k^{\ast}}{||\bm h_k||}|^2 + \sigma^2$. Thus the information rate for the $0$-th D2D Rx (for fixed channel gain) is given by $\log_2(1+ \text{SINR})$, where $\text{SINR} \Define \gamma_d\beta_0||\bm h_0||^2/(\gamma_d\sum_{k \in \Phi}\beta_k^{(d)}|\frac{\bm h_k^{(d), \, T}\, \bm h_k^{\ast}}{||\bm h_k||}\,|^2 + 1)$. Here $\gamma_d \Define P_D/\sigma^2$. The ergodic ASE is therefore given by $\lambda\E[ \log_2 (1 + \text{SINR})] \Define R^{(d)}$, i.e.,

\vspace{-0.5 cm}

\begin{IEEEeqnarray}{rCl}
\label{eq:ergo_ase}
R^{(d)} & = & \lambda \E \Bigg[\log_2 \Big(1 + \frac{\gamma_d \beta_0 ||\bm h_0||^2}{\gamma_d \sum_{k \in \Phi} \beta_k^{(d)}\Big|\frac{\bm h_k^{(d)T} \bm h_k^{\ast}}{||\bm h_k||}\Big|^2 + 1}\Big)\Bigg].
\IEEEeqnarraynumspace
\end{IEEEeqnarray}

\vspace{-0.2 cm}

\indent Here the expectation is taken over all realizations of channel gain vectors and over all realizations of PPP $\Phi$.

\begin{proposition}
\label{aselb}
(Lower Bound on ASE)
\normalfont
A lower bound on the ergodic ASE of the underlaid D2D network, defined in \eqref{eq:ergo_ase}, is given by

\vspace{-0.7 cm}

\begin{IEEEeqnarray}{rCl}
\label{eq:d2dratelb}
R^{(d)} \geq R_{\lambda}^{(d)} & = & \lambda \, \log_2 \Big(1 + \frac{(N - 1) (d_0/D)^{\alpha_d} \, \gamma_d}{c_1 \big[1 + \frac{\lambda  \pi d_0^2 \alpha_d c_0 \gamma_d}{\alpha_d - 2}\big]}\Big),
\IEEEeqnarraynumspace
\end{IEEEeqnarray}

\vspace{-0.15 cm}

\noindent where $c_0 \Define \left(\frac{\lambda_c}{4\pi d_0}\right)^2 e^{\frac{\sigma_{\text{db}}^2}{200} (\ln10)^2}$, $c_1 \Define \left(\frac{4\pi d_0}{\lambda_c}\right)^2 e^{\frac{\sigma_{\text{db}}^2}{200} (\ln10)^2}$.
\end{proposition}

\begin{IEEEproof}
See Appendix. 
\end{IEEEproof}

\begin{remark}
\normalfont
It is observed that with fixed $N$, $\lambda$ and $D$ and with $P_D \to \infty$, i.e., $\gamma_d \to \infty$, we have

\vspace{-0.6 cm}

\begin{IEEEeqnarray}{rCl}
\label{eq:limase}
R_{\lambda}^{\infty} & \Define & \lim\limits_{\gamma_d \to \infty} R_{\lambda}^{(d)} =  \lambda \log_2 \Bigg(1 + \frac{(N - 1)(1 - {2}/{\alpha_d})}{\lambda \pi d_0^2 c_0 c_1 ({d_0}/{D})^{-\alpha_d}} \Bigg),
\IEEEeqnarraynumspace
\end{IEEEeqnarray}

\vspace{-0.2 cm}

\noindent i.e., the achievable ASE approaches a constant limiting value. The ASE, $R_{\lambda}^{(d)}$, is a monotonically increasing function of the D2D Tx power, $P_D$ (see \eqref{eq:d2dratelb}). Therefore with increasing $\gamma_d = P_D/\sigma^2$, both the useful received signal power at the D2D Rx under consideration as well as the total interference power from all other interfering D2D Txs increase linearly (since the AWGN power, $\sigma^2$, remains fixed). Therefore the D2D ASE becomes interference limited as $\gamma_d \to \infty$. Hence it is clear from \eqref{eq:limase} that for a given area density of D2D Txs, $\lambda$, the achievable ASE, $R_{\lambda}^{(d)}$ cannot exceed the corresponding limiting value, $R_{\lambda}^{\infty}$, for that given $\lambda$, i.e., $R_{\lambda}^{(d)} < R_{\lambda}^{\infty}$. \hfill\qed
\end{remark}

\begin{remark}
\label{toasevslamda}
\normalfont
From \eqref{eq:d2dratelb}, using $R_{\lambda}^{(d)} = R_0^{(d)}$ and fixed $N$ and $D$, we have

\vspace{-0.6 cm}

\begin{IEEEeqnarray}{rCl}
\label{eq:intf1}
\lambda \, \gamma_d & = & \Bigg[\frac{(\frac{N - 1}{c_1})( {d_0}/{D})^{\alpha_d}}{\lambda (2^{R_{0}^{(d)}/\lambda} - 1)} - \frac{\pi d_0^2 \alpha_d c_0}{(\alpha_d - 2)}\Bigg]^{-1}.
\IEEEeqnarraynumspace
\end{IEEEeqnarray}

\vspace{-0.2 cm}

\indent Since $\lambda (2^{R_0^{(d)}/\lambda}-1)$ is monotonically decreasing with increasing $\lambda$ for a fixed $ R_0^{(d)}$, it is evident from \eqref{eq:intf1} that the \textit{total interference power at the D2D Rx under consideration ($\propto \lambda \gamma_d$) decreases with increasing $\lambda$}. Further the useful signal power received at the D2D Rx under consideration is proportional to $\gamma_d$. Thus we have 

\vspace{-0.6 cm}

\begin{IEEEeqnarray}{rCl}
\label{eq:sinreq2}
\text{SINR}_{\text{req}} &= & {c_2 \, \gamma_d}/{(1 + c_3 \, \lambda \gamma_d)},
\IEEEeqnarraynumspace
\end{IEEEeqnarray}

\vspace{-0.2 cm}

\noindent where $c_2>0$ and $c_3>0$ are constants. For a fixed desired D2D ASE $R_0^{(d)}$, the average per-user spectral efficiency (SE) is given by $r^{(d)} \Define R_0^{(d)}/\lambda$. For small enough SE ($r^{(d)} \ll 1$), we know that the required SINR is proportional to the SE, i.e., $r^{(d)} = \log_2(1+\text{SINR}) \implies \text{SINR} = 2^{r^{(d)}} - 1 \approx r^{(d)} \ln 2$. Thus for a fixed $R_0^{(d)}$ and increasing $\lambda$, we have $\text{SINR}_{\text{req}} = 2^{R_{0}^{(d)}/\lambda} - 1 \approx ({R_{0}^{(d)}}/{\lambda}) \ln 2$, when $\lambda \gg R_0^{(d)}$, i.e. when $\lambda$ is sufficiently large. Using $\text{SINR}_{\text{req}} \approx ({R_{0}^{(d)}}/{\lambda}) \ln 2$ in \eqref{eq:sinreq2}, we get $\frac{c_2 \, \lambda \, \gamma_d}{1 + c_3 \, \lambda \, \gamma_d} = R_{0}^{(d)} \, \ln 2$. Thus for a fixed desired D2D ASE $R_0^{(d)}$, \textit{when $\lambda$ is sufficiently large}, i.e., $\lambda \gg R_0^{(d)}$, we have\footnote[4]{This can also be shown from \eqref{eq:intf1} using the approximation $e^x \approx 1+x$ (for small $x$) on $\lambda(2^{R_0^{(d)}/\lambda}-1)$.} $\lambda \gamma_d = $ constant, or, $\gamma_d \propto \frac{1}{\lambda}$. \hfill \qed
\end{remark}

The results in Remark~\ref{aselb} and Remark~\ref{toasevslamda} would be later used in the derivation of the main result in this paper (Theorem~\ref{cellratease1} in Section III).

\vspace{-0.5 cm}

\section{Average Per-User Spectral Efficiency of Cellular Downlink}

For cellular downlink in massive MIMO, the per-user spectral efficiency (SE) of CUEs in the absence of underlaid D2D interference can be modelled as

\vspace{-0.7 cm}

\begin{IEEEeqnarray}{rCl}
\label{eq:cellsenod2d}
R_{0}^{(c)} = \log_2(1 + \gamma_{0}^{c}),
\IEEEeqnarraynumspace
\end{IEEEeqnarray}

\vspace{-0.2 cm}

\noindent where $\gamma_0^c$ is the effective average SINR at the CUEs. In the absence of D2D interference, the downlink SINR at the CUE, $\gamma_{0}^{c}$ is almost deterministic (i.e. does not vary significantly with changing channel realization) when the number of BS antennas is sufficiently large \cite{Marzetta1}. For a given density of D2D Txs, $\lambda$, let $I_d$ be the total interference power from the D2D Txs to the CUE under consideration at the origin. The channel gain vector between the CUE at origin and the $l^{\text{th}}$ D2D Tx is given by $\bm g_l^{(c)} = (g_{l,1}^{(c)}, g_{l,2}^{(c)}, \cdots, g_{l,N}^{(c)})^T = \sqrt{\beta_l^{(c)}} \bm h_l^{(c)} \in \C^{N \times 1}$, where $g_{l,i}^{(c)}, i = 1, 2, \ldots, N$, is the channel gain between the $i^{\text{th}}$ antenna of the $l^{\text{th}}$ D2D Tx and the single antenna of the CUE under consideration. $\bm h_l^{(c)} \sim \mathcal{C}\mathcal{N}(0, \bm I_N)$ are i.i.d. and $\sqrt{\beta_l^{(c)}} > 0$ models the corresponding geometric attenuation. With MRT beamforming from the $l$-th D2D Tx to its corresponding D2D Rx, the overall interfering D2D signal received at the CUE at the origin is given by $s_d  = \sqrt{P_D} \sum_{l \in \Phi} \frac{\bm g_l^{(c)T}\bm g_l^{\ast}}{||\bm g_l||}u_l$, where $\bm g_l = \sqrt{\beta_l}\, \bm h_l \in \C^{N \times 1}$ is the channel gain vector between the $l^{\text{th}}$ D2D Tx-Rx pair and $u_l \sim \mathcal{C}\mathcal{N}(0,1)$ is the information symbol transmitted by the $l^{\text{th}}$ D2D Tx. Note that the D2D interference signal is Gaussian for a given channel realization. Clearly the total D2D interference power at the CUE under consideration is given by

\vspace{-0.6 cm}

\begin{IEEEeqnarray}{rCl}
\label{eq:interfd2d}
I_d \Define \E[|s_d|^2] = P_D \sum\limits_{l \in \Phi} \beta_l^{(c)} \Big|\frac{\bm h_l^{(c)T}\bm h_l^{\ast}}{||\bm h_l||}\Big|^2.
\IEEEeqnarraynumspace
\end{IEEEeqnarray}

\vspace{-0.2 cm}

\indent The power of the information signal received at the CUE under consideration from the BS is $\gamma_{0}^{(c)} \, \sigma^2$ and the AWGN power at the CUE is $\sigma^2$ (see \eqref{eq:cellsenod2d}). Therefore in the presence of D2D interference, the total effective SINR for a given channel realization is given by $\gamma_{0}^{(c)} \sigma^2 /(I_d + \sigma^2) = \gamma_{0}^{(c)}/(I_d/\sigma^2 + 1)$. The ergodic per-user SE of the CUE under consideration is therefore given by

\vspace{-0.6 cm}

\begin{IEEEeqnarray}{rCl}
\label{eq:celld2d}
R^{c,d} & = & \E \Big[\log_2 \Big( 1 + \frac{\gamma_{0}^c}{{I_d}/{\sigma^2} + 1} \Big)\Big],
\IEEEeqnarraynumspace
\end{IEEEeqnarray}

\vspace{-0.2 cm}

\noindent where $\E[.]$ indicates averaging over all possible realizations of channel gain vectors and also over all realizations of PPP $\Phi$.\footnote[5]{For the derivation of the per-user cellular rate, we consider $\gamma_0^c$ to be fixed and deterministic as discussed above.}

\begin{proposition}
\label{cellratelbd2d}
(Average Per-User spectral efficiency (SE) for CUE in the presence of D2D Interference)
\normalfont
An average per-user SE of CUEs in the presence of D2D interference is given by

\vspace{-0.6 cm}

\begin{IEEEeqnarray}{rCl}
\label{eq:celld2dlb}
R_{\lambda}^{c,d} & = & \log_2\Big(1 + \frac{2^{R_{0}^{(c)}} - 1}{1 + \vartheta \, \lambda \, \gamma_d}\Big),
\IEEEeqnarraynumspace
\end{IEEEeqnarray}

\vspace{-0.2 cm}

\noindent where $\gamma_d = P_D/\sigma^2$ and $\vartheta \Define \frac{\pi d_0^2 \, c_0 \, \alpha_d}{\alpha_d - 2}$.
\end{proposition}

\begin{IEEEproof}
Using Jensen's inequality in \eqref{eq:celld2d}, we have

\vspace{-0.55 cm}

\begin{IEEEeqnarray}{rCl}
\label{eq:cellratelb}
R^{c,d} \myageq \log_2 \Big(1 + \dfrac{2^{R_{0}^{(c)}} - 1}{ 1 + \E[I_d]/\sigma^2}\Big) \Define R_{\lambda}^{c,d},
\IEEEeqnarraynumspace
\end{IEEEeqnarray}

\vspace{-0.2 cm}

\noindent where $(a)$ follows from \eqref{eq:cellsenod2d}. Using the expression of $I_d$ from \eqref{eq:interfd2d}, we have

\vspace{-0.6 cm}

{\begin{IEEEeqnarray}{rCl}
\label{eq:intfpavg}
\nonumber \E[I_d] & \myb & P_D \E\Big[\sum\limits_{l \in \Phi} \beta_l^{(c)}\Big] = P_D \E_{\Phi}\Big[\sum\limits_{l \in \Phi}\E[\beta_l^{(c)}]\Big]\\
\nonumber & \myc & P_D \, c_0 \,  \E_{\Phi}\Big[\sum\limits_{l \in \Phi} \dfrac{1}{l(x_l)}\Big] \, \myd \, P_D \,  c_0 \int\limits_{0}^{\infty} \frac{1}{l(x)} 2\lambda \pi x dx\\
 & \mye &  \big(\frac{\pi d_0^2 \, c_0 \, \alpha_d}{\alpha_d - 2}\big) \lambda \, P_D = \vartheta \, \lambda \, P_D,
\IEEEeqnarraynumspace
\end{IEEEeqnarray}}

\vspace{-0.5 cm}

\noindent where $(b)$ follows from the fact that $\E_{\bm h_l^{(c)}}\Big[ \big|\frac{\bm h_l^{(c)T}\bm h_l^{\ast}}{\| \bm h_l \|}\big|^2 \Big|\bm h_l\Big] = 1$. Step $(c)$ follows from the fact that $\E[\beta_l^{(c)}] = c_0/l(x_l)$, where $x_l$ is the distance of the $l^{\text{th}}$ D2D Tx from the CUE under consideration at the origin.\footnote[6]{The result $\E[\beta_l^{(c)}] = c_0/l(x_l)$ is obtained by proceeding along the same steps as in \eqref{eq:avgloss} (see Appendix).} Step $(d)$ follows from Campbell's theorem \cite{Haenggi}. Finally step $(e)$ follows from the definition of $l(x)$ (see the line after \eqref{eq:pathloss2}). Using \eqref{eq:intfpavg} in \eqref{eq:cellratelb}, we obtain \eqref{eq:celld2dlb}.
\end{IEEEproof}

In the following using Propositions~\ref{aselb} and ~\ref{cellratelbd2d}, we derive the main result of this paper where we characterize the per-user SE of CUE as a function of D2D ASE.

\begin{theorem}
\label{cellratease1}
For a fixed desired ASE of the D2D network $R_{\lambda}^{(d)} = R_{0}^{(d)} < R_{\lambda}^{\infty}$ ($R_{\lambda}^{\infty}$ is defined in \eqref{eq:limase}), the average per-user spectral efficiency of cellular downlink in the presence of D2D interference is given by

\vspace{-0.6 cm}

\begin{IEEEeqnarray}{rCl}
\label{eq:cellase}
R_{\lambda}^{c,d} & = & \log_2 (1 + (2^{R_{0}^{(c)}} - 1) (1 - \kappa)),
\IEEEeqnarraynumspace
\end{IEEEeqnarray}

\vspace{-0.2 cm}

\noindent where $\kappa \Define \frac{\lambda (2^{R_{0}^{(d)}/\lambda} - 1)}{N - 1}\Theta$ and $\Theta \Define \frac{ \pi d_0^2 c_0 c_1\alpha_d}{(d_0/D)^{\alpha_d} (\alpha_d - 2)}$.
\end{theorem}

\begin{IEEEproof}
Since $R_{0}^{(d)} < R_{\lambda}^{\infty}$, the desired ASE is achievable for the given $\lambda$ (see Remark~\ref{aselb}). $R_{\lambda}^{c,d}$ depends on $\gamma_d$ (see \eqref{eq:celld2dlb} in Proposition~\ref{cellratelbd2d}) only through the total D2D interference (see the term $\lambda \gamma_d$ in the denominator of the R.H.S. of \eqref{eq:celld2dlb}). For a fixed D2D ASE $R_{0}^{(d)}$, we get the expression for the required $\gamma_d$ (in terms of $R_{0}^{(d)}$) from \eqref{eq:intf1} in Remark~\ref{toasevslamda}. Substituting this expression of $\gamma_d$ in \eqref{eq:celld2dlb} gives us \eqref{eq:cellase}.
\end{IEEEproof}

\begin{figure}[t]
\vspace{-0.5 cm}
\includegraphics[width= 3.4 in, height= 1.8 in]{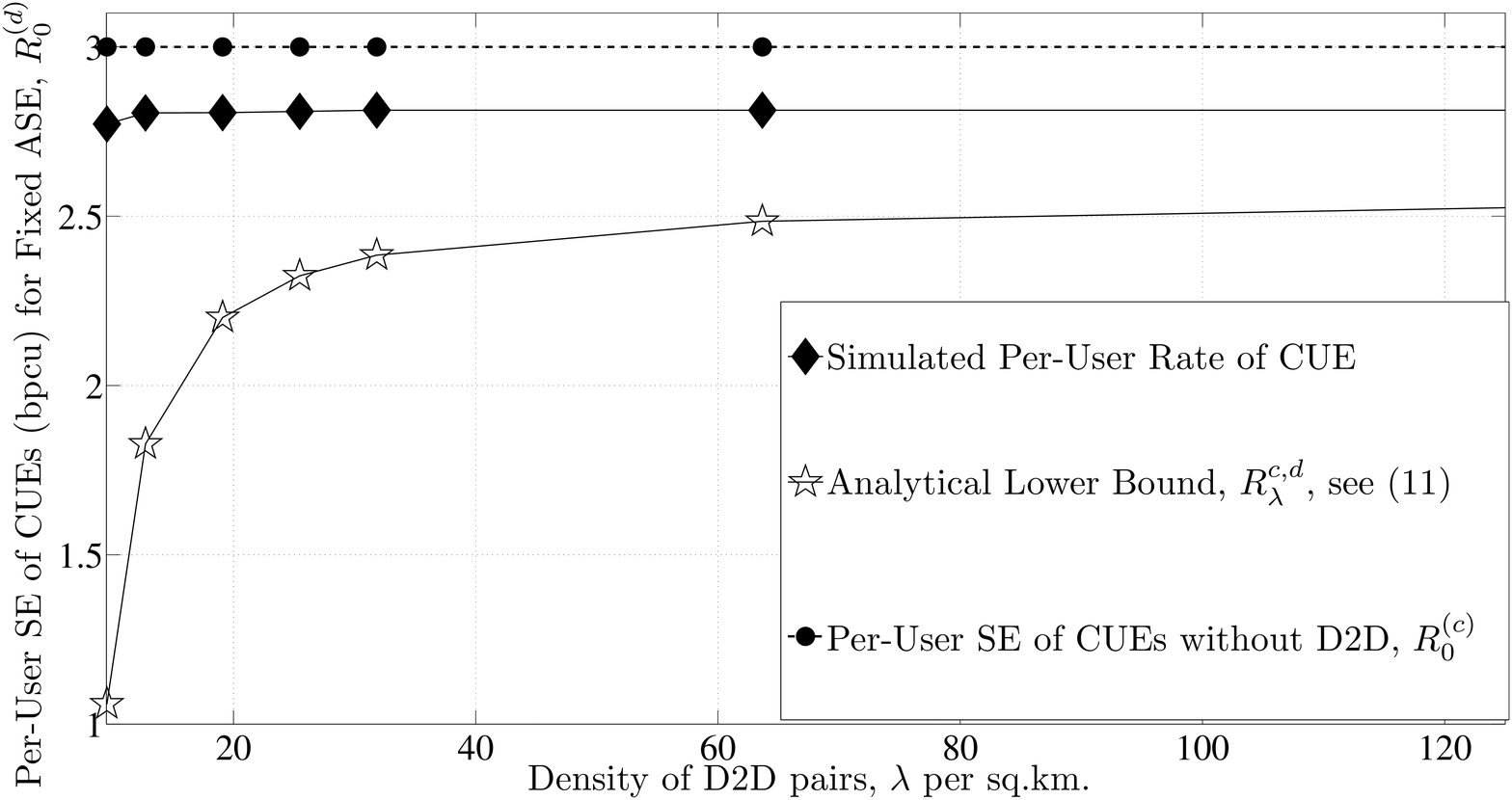}
\caption {Plot of the achievable average per-user SE of CUEs as a function of increasing D2D pair density $\lambda$ for a fixed desired D2D ASE $R_{0}^{(d)} = 25$ bps/Hz/sq.km. and the following fixed parameters: $N = 4$, $D = 50$m, $\alpha_d = 4$, $d_0 = 35$m, standard deviation of lognormal shadowing $\sigma_{\text{db}} = 3$ dB and noise variance $\sigma^2 = -114$ dBm.}
\vspace{-0.5 cm}
\label{fig:figcellvsase}
\end{figure}

\begin{remark}
\label{cellratease}
\normalfont
In the absence of any D2D interference, the average per-user SE of CUE is $R_{0}^{(c)}$. With a few D2D pairs, the interference increases from zero to some finite value. This results in decrease in the SE of CUE. With further increase in the number of D2D pairs (i.e. increasing $\lambda$), for a fixed desired ASE of the D2D network, interestingly, the total D2D interference power ($\propto \lambda P_D$) decreases (see Remark~\ref{toasevslamda}). \textit{This results in an increase in the per-user SE of CUEs with increasing $\lambda$ and fixed D2D ASE} (see also Fig.~\ref{fig:figcellvsase}). \hfill \qed
\end{remark}

Further, when $\lambda$ is sufficiently large (with fixed D2D ASE), from Remark~\ref{toasevslamda} we also know that the total D2D Tx interference power ($\propto \lambda P_D$) approaches a constant limiting value. \textit{This suggests that with fixed D2D ASE and $\lambda \to \infty$, the per-user SE of CUEs also approaches a  limiting value less than $R_{0}^{(c)}$}. Thus from \eqref{eq:cellase} we have $R_{\infty}^{c,d} \Define \lim\limits_{\lambda \to \infty} R_{\lambda}^{c,d}$, i.e,

\vspace{-0.45 cm}

\begin{IEEEeqnarray}{rCl}
R_{\infty}^{c,d} & = & \log_2 \Big(1 + (2^{R_{0}^{(c)}} - 1) (1 - {\Theta{R_{0}^{(d)}} \, \ln 2}/{(N - 1)})\Big).
\label{eq:limitcellase}
\IEEEeqnarraynumspace
\end{IEEEeqnarray}

\vspace{-0.2 cm}

\indent This value of $R_{\infty}^{c,d}$ gives us a \textit{fundamental limit on the average per-user SE of CUEs} that can be achieved while maintaining a fixed desired D2D ASE. For a fixed D2D ASE, $R_{0}^{(d)}$, the function $g(\lambda) \Define \lambda(2^{R_{0}^{(d)}/\lambda} - 1)$ decreases very fast initially with increasing $\lambda$ and then decreases slowly to its limiting value $R_{0}^{(d)} \ln 2$. This behaviour of $g(\lambda)$ in the numerator of $\kappa$ (see \eqref{eq:cellase}) \textit{suggests that beyond a certain value of $\lambda$, the increase in the average per-user SE of CUEs is small}. This is also observed in Fig.~\ref{fig:figcellvsase}.

\begin{remark}
\label{remNvsase}
\normalfont
In \eqref{eq:limitcellase} we note that for a fixed D2D ASE $R_{0}^{(d)}$ and fixed number of D2D Tx antennas, $N$, the fundamental limit on the average per-user SE of CUEs ($R_{\infty}^{c,d}$) depends on $R_{0}^{(d)}$ and $N$ \textit{only through the ratio} $\frac{R_{0}^{(d)}}{N - 1}$. This means that a larger number of D2D Tx antennas ($N$) would allow us to achieve the same $R_{\infty}^{c,d}$ for a higher value of D2D ASE ($R_{0}^{(d)}$). This suggests that with every doubling in $N$, the D2D ASE can be approximately doubled. This observation is supported by Fig.~\ref{fig:figasevsN}. \hfill \qed
\end{remark}

\vspace{-0.5 cm}

\section{Numerical Results and Discussions}

In this section, we use monte-carlo simulation to verify the relation between the average per-user SE of CUEs and the D2D ASE. For all simulation studies, we assume the following values for system parameters: operating carrier frequency $f_c = 2$ GHz, the distance between Tx and Rx of a D2D pair $D = 50$m and the far-field distance from transmitter $d_0 = 35$m (for outdoor environment in a micro-cell)\cite{Goldsmith,Bai}. Further we assume the noise power spectral density to be $N_0 = 10^{-20.4}$ W/Hz and the communication bandwidth to be $B_{\text{w}} = 1$ MHz. Thus the effective noise power is given by $\sigma^2 = N_0 B_{\text{w}} = -114$ dBm. The pathloss exponent is taken to be $\alpha_d = 4$ and the standard deviation for lognormal shadow fading is $\sigma_{\text{db}} = 3$ dB. All these values are based on realistic data obtained in prior works \cite{Lin, Lte, Shalmashi}. Also we assume the average per-user SE of CUEs in massive MIMO downlink in the absence of underlaid D2D to be $R_{0}^{(c)} = 3$ bps/Hz.

\par In Fig.~\ref{fig:figcellvsase}, for a fixed desired D2D ASE $R_{0}^{(d)} = 25$ bps/Hz/sq.km. and fixed $N = 4$, we plot the exact average per-user SE of CUEs (by numerically computing the expectation in \eqref{eq:celld2d}) as a function of increasing D2D pair density $\lambda$. We also plot the analytical lower bound to \eqref{eq:celld2d}, i.e. the achievable rate $R_{\lambda}^{c,d}$ (see \eqref{eq:celld2dlb}). It is observed that for sufficiently large $\lambda$, the average per-user SE of CUEs increases monotonically and approaches a limiting value. This supports our conclusions regarding the fundamental limit on the average per-user SE of CUEs in Remark~\ref{cellratease}. From the figure we note that the difference in the values of this fundamental limit obtained from simulation and that from our analytical expression is small ($\approx 0.2$ bps/Hz).

\par
\begin{figure}[t]
\vspace{-0.5 cm}
\includegraphics[width= 3.4 in, height= 1.8 in]{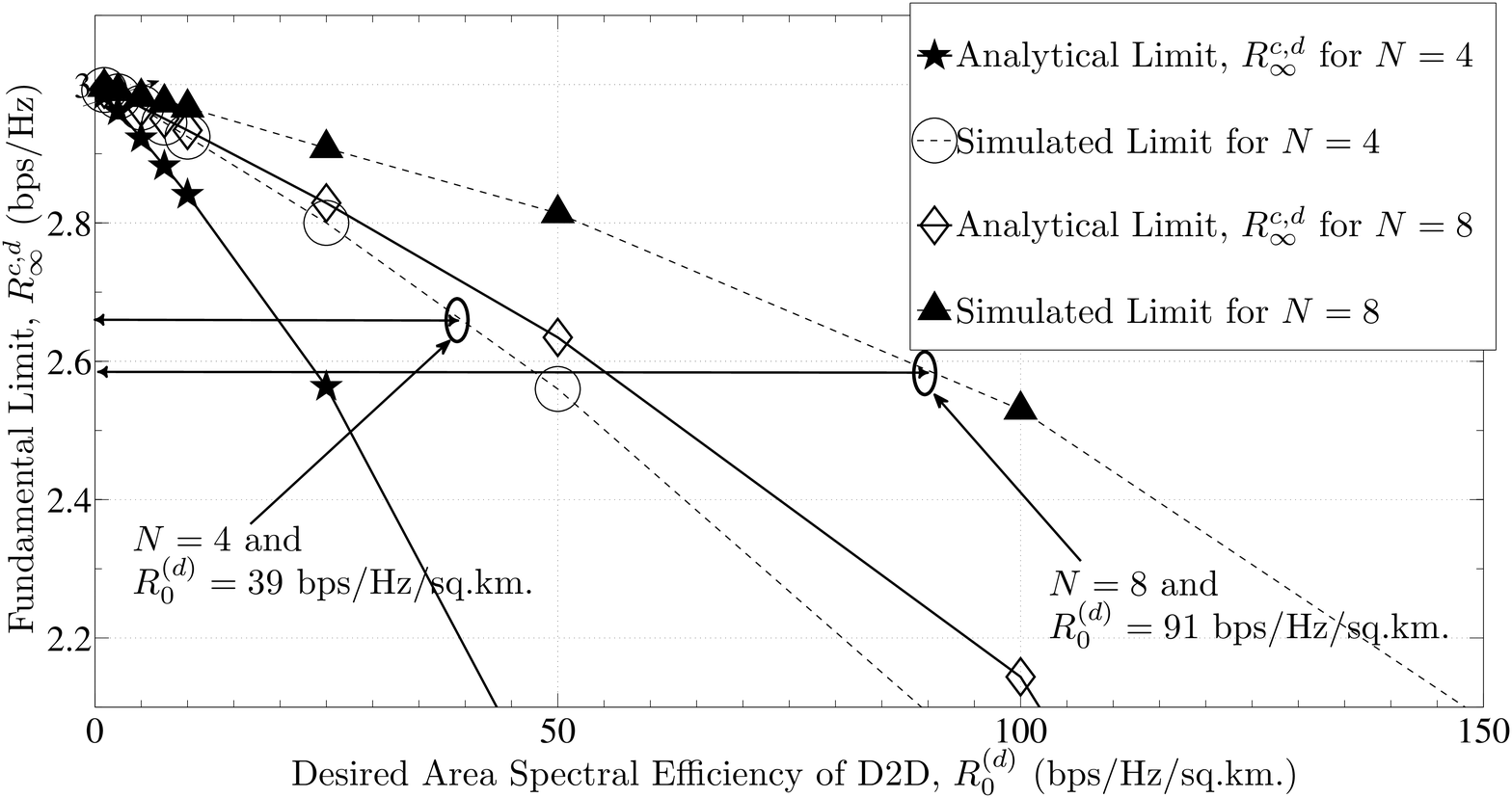}
\caption {Plot of the fundamental limit on the average per-user SE of CUEs as a function of increasing D2D ASE $R_{0}^{(d)}$ for $N = 4$ and $N = 8$. The fixed parameters are: $D = 50$m, $\alpha_d = 4$, $d_0 = 35$m, noise variance $\sigma^2 = -114$ dBm and standard deviation of lognormal shadowing $\sigma_{\text{db}} = 3$ dB.}
\vspace{-0.5 cm}
\label{fig:figasevsN}
\end{figure}

 In Fig.~\ref{fig:figasevsN}, we plot the fundamental limit on the average per-user SE of CUEs, both analytically (see \eqref{eq:limitcellase}) and numerically through simulation, as a function of increasing D2D ASE, $R_{0}^{(d)}$ for fixed $N = 4, 8$. It is observed that with increasing $R_{0}^{(d)}$, the fundamental limit on the per-user SE of CUEs decreases monotonically. From the figure it is also clear that for a fixed $\frac{R_{0}^{(d)}}{N - 1}$, the fundamental limit on the per-user SE of CUEs remains almost constant. For instance, in the figure, with $N = 4$ and $R_{0}^{(d)} = 39$ bps/Hz/sq.km. (see the dashed curve with circles) we have $R_{\infty}^{c,d} \approx 2.66$ bps/Hz and with $N = 8$ and $R_{0}^{(d)} = 91$ bps/Hz/sq.km. (see the dashed curve with filled triangles), we have $R_{\infty}^{c,d} \approx 2.59$ bps/Hz. Note that in both cases $\frac{R_{0}^{(d)}}{N - 1} = 13$, $R_{\infty}^{c,d}$ is almost the same and the D2D ASE almost doubles ($91/39 = 2.33$). This supports our conclusion in Remark~\ref{remNvsase}.

\vspace{-0.5 cm}

\section{Conclusion}

Thus our work presents two important conclusions on the impact of D2D interference on the average per-user information rate of CUEs in massive MIMO downlink systems. Firstly, through Remark~\ref{cellratease} we show that for a fixed D2D ASE, with sufficiently large density of D2D pairs, the average per-user SE of CUEs increases monotonically with increasing density of D2D pairs and approaches a limit. Secondly, in Remark~\ref{remNvsase} we show that a larger number of D2D Tx antennas would allow a higher value of D2D ASE to be achieved while maintaining a fixed fundamental limit on the average per-user SE of CUEs.

\vspace{-0.5 cm}

\appendix[Proof of Proposition 1]

Using Jensen's inequality in \eqref{eq:ergo_ase}, we have

\vspace{-0.55 cm}

\begin{IEEEeqnarray}{rCl}
\label{eq:aselb2}
R^{(d)} & \geq & \lambda \log_2 \Big(1 + \frac{1}{\E\Big[{1}/{\text{SINR}}\Big]}\Big) \Define R_{\lambda}^{(d)},
\IEEEeqnarraynumspace
\end{IEEEeqnarray}

\vspace{-0.2 cm}

 \noindent where $\text{SINR} = \gamma_d\beta_0||\bm h_0||^2/(\gamma_d\sum_{k \in \Phi}\beta_k^{(d)}|\frac{\bm h_k^{(d), \, T}\, \bm h_k^{\ast}}{||\bm h_k||}\,|^2 + 1)$. Therefore we have 

\vspace{-0.5 cm}

\small{\begin{IEEEeqnarray}{rCl}
\label{eq:avginvsinr}
\E\left[\frac{1}{\text{SINR}}\right] & \mya & \E\Big[\frac{1/\gamma_d}{\beta_0 ||\bm h_0||^2}\Big] \E\Bigg[\gamma_d \sum_{k \in \Phi} \beta_k^{(d)} \Big|\frac{\bm h_k^{(d)T} \bm h_k^{\ast}}{||\bm h_k||}\Big|^2 + 1\Bigg],
\IEEEeqnarraynumspace
\end{IEEEeqnarray}}\normalsize

\vspace{-0.2 cm}

\noindent where $(a)$ follows from the fact that $\beta_0$, $\beta_k$, $\beta_k^{(d)}$, $\bm h_k$, $\bm h_{k}^{(d)}$ and $\bm h_{0}$ are all independent. Further, $||\bm h_0||^2 \sim \chi ^2 (2N)$ distributed and therefore $\E\big[\frac{1}{||\bm h_0||^{2}}\big] = {1}/{(N - 1)}$. From \eqref{eq:pathloss2} for a given realization of PPP $\Phi$ we have

\vspace{-0.4 cm}

\begin{IEEEeqnarray}{rCl}
\label{eq:avgloss}
\E[\, \beta_{k}^{(d)} \, ] & \myb & \left({\lambda_c}/{4\pi d_0}\right)^2 \E\Big[e^{-\frac{\Xi_{\text{dB}}}{10} \ln 10}\Big]/{l(r_k)} = {c_0}/{l(r_k)},
\IEEEeqnarraynumspace
\end{IEEEeqnarray}

\vspace{-0.2 cm}

\noindent where $ c_0 = \left(\frac{\lambda_c}{4\pi d_0}\right)^2 e^{\frac{\sigma_{\text{db}}^2}{200} (\ln10)^2}$ and step $(b)$ follows from the fact that $\Xi_{\text{dB}} \sim \mathcal{N}(0,\sigma_{\text{db}}^2)$. In a similar fashion, it can also be shown that 

\vspace{-0.6 cm}

\begin{IEEEeqnarray}{rCl}
\label{eq:c1}
\E[1/\beta_0] = c_1 l(D) = c_1 (D/d_0)^{\alpha_d},
\IEEEeqnarraynumspace
\end{IEEEeqnarray}

\vspace{-0.2 cm}

\noindent where $c_1 \Define \left({4\pi d_0}/{\lambda_c}\right)^2 \E\Big[e^{\frac{\Xi_{\text{dB}}}{10} \ln 10}\Big]  = \left(\frac{4\pi d_0}{\lambda_c}\right)^2 e^{\frac{\sigma_{\text{db}}^2}{200} (\ln 10)^2}$.

\par Further it can be shown that when conditioned on $\bm h_k$, $\frac{\bm h_k^{(d)T} \bm h_k^{\ast}}{||\bm h_k||}$ is complex Gaussian zero mean with variance $1$. Using this fact, the second term in the R.H.S. of \eqref{eq:avginvsinr} can be simplified as 

\vspace{-0.5 cm}

{\begin{IEEEeqnarray}{lCl}
\label{eq:intfp}
\nonumber \E\left[\gamma_d \sum\limits_{k \in \Phi} \beta_k^{(d)} \left|\frac{\bm h_k^{(d)T} \bm h_k^{\ast}}{||\bm h_k||}\right|^2 + 1\right]  = \E_{\Phi}\left[\gamma_d \sum\limits_{k \in \Phi} \E\Big[\beta_k^{(d)}\Big] \right] + 1\\
\nonumber \myc  \gamma_d \, c_0 \E_{\Phi}\left[\sum\limits_{k \in \Phi} \frac{1}{l(r_k)}\right] + 1  \myd  \gamma_d \, c_0 \int\limits_{0}^{\infty}\frac{1}{l(r)} 2\lambda \pi r dr + 1\\
 \mye  \dfrac{\lambda \pi d_0^2 \, \alpha_d \, c_0 \, \gamma_d}{\alpha_d - 2} + 1,
\IEEEeqnarraynumspace
\end{IEEEeqnarray}}

\vspace{-0.4 cm}

\noindent where step $(c)$ follows from \eqref{eq:avgloss} and step $(d)$ follows from Campbell's theorem \cite{Haenggi}. Step $(e)$ follows from the definition of $l(r)$ (see the line after \eqref{eq:pathloss2}). Substituting \eqref{eq:intfp} and \eqref{eq:c1} in \eqref{eq:avginvsinr} we have

\vspace{-0.6 cm}

\begin{IEEEeqnarray}{rCl}
\label{eq:invsinr}
\E\left[\frac{1}{\text{SINR}}\right] & = & \frac{c_1 \big[1 + \frac{\lambda \pi d_0^2 \alpha_d c_0 \gamma_d}{\alpha_d - 2}\big]}{(N - 1)(d_0/D)^{\alpha_d} \gamma_d}.
\IEEEeqnarraynumspace
\end{IEEEeqnarray}

\vspace{-0.15 cm}

\noindent Substituting \eqref{eq:invsinr} in the expression of $R_{\lambda}^{(d)}$ in \eqref{eq:aselb2}, we get \eqref{eq:d2dratelb}.

\vspace{-0.35 cm}



%

%


\ifCLASSOPTIONcaptionsoff
  \newpage
\fi



%

\bibliographystyle{IEEEtran}
\bibliography{IEEEabrvn,mybibn}

\begin{thebibliography}{10}
\providecommand{\url}[1]{#1}
\csname url@samestyle\endcsname
\providecommand{\newblock}{\relax}
\providecommand{\bibinfo}[2]{#2}
\providecommand{\BIBentrySTDinterwordspacing}{\spaceskip=0pt\relax}
\providecommand{\BIBentryALTinterwordstretchfactor}{4}
\providecommand{\BIBentryALTinterwordspacing}{\spaceskip=\fontdimen2\font plus
\BIBentryALTinterwordstretchfactor\fontdimen3\font minus
  \fontdimen4\font\relax}
\providecommand{\BIBforeignlanguage}[2]{{%
\expandafter\ifx\csname l@#1\endcsname\relax
\typeout{** WARNING: IEEEtran.bst: No hyphenation pattern has been}%
\typeout{** loaded for the language `#1'. Using the pattern for}%
\typeout{** the default language instead.}%
\else
\language=\csname l@#1\endcsname
\fi
#2}}
\providecommand{\BIBdecl}{\relax}
\BIBdecl

\bibitem{Lte}
\lq\lq{3rd Generation Partnership Project: Technical Specification Group Radio
  Access Network: Study on LTE device to device proximity services; Radio
  aspects (Release 12)}", Cedex, France, TR 36.843 V12.0.1, March 2014.

\bibitem{Lin2}
X.~Lin, J.~G. Andrews, A.~Ghosh, and R.~Ratasuk, ``An {O}verview of {3GPP}
  {D}evice-to-{D}evice {P}roximity {S}ervices,'' \emph{{IEEE} Commun. Mag.},
  vol.~52, no.~4, pp. 40--48, April 2014.

\bibitem{Tehrani}
M.~Tehrani, M.~Uysal, and H.~Yanikomeroglu, ``Device-to-{D}evice
  {C}ommunication in 5{G} {C}ellular {N}etworks: {C}hallenges, {S}olutions, and
  {F}uture {D}irections,'' \emph{{IEEE} Commun. Mag.}, vol.~52, no.~5, pp.
  86--92, May 2014.

\bibitem{Shalmashi}
S.~Shalmashi, E.~Bjornson, M.~Kountouris, K.~W. Sung, and M.~Debbah, ``Energy
  {E}fficiency and {S}um {R}ate when {M}assive {MIMO} meets
  {D}evice-to-{D}evice {C}ommunication,'' in \emph{Communication Workshop
  (ICCW), 2015 IEEE International Conference on}, June 2015, pp. 627--632.

\bibitem{Andrews}
J.~Andrews, S.~Buzzi, W.~Choi, S.~Hanly, A.~Lozano, A.~Soong, and J.~Zhang,
  ``What {W}ill 5{G} {B}e?'' \emph{{IEEE} J. Sel. Areas Commun.}, vol.~32,
  no.~6, pp. 1065--1082, June 2014.

\bibitem{Marzetta2}
F.~Rusek, D.~Persson, B.~K. Lau, E.~Larsson, T.~Marzetta, O.~Edfors, and
  F.~Tufvesson, ``Scaling {U}p {MIMO}: {O}pportunities and {C}hallenges with
  {V}ery {L}arge {A}rrays,'' \emph{{IEEE} Signal Process. Mag.}, vol.~30,
  no.~1, pp. 40--60, Jan 2013.

\bibitem{Lin}
X.~Lin, R.~W. Heath, and J.~G. Andrews, ``The {I}nterplay {B}etween {M}assive
  {MIMO} and {U}nderlaid {D2D} {N}etworking,'' \emph{{IEEE} Trans. Wireless
  Commun.}, vol.~14, no.~6, pp. 3337--3351, June 2015.

\bibitem{Goldsmith}
A.~J. Goldsmith, \emph{Wireless {C}ommunications}.\hskip 1em plus 0.5em minus
  0.4em\relax Cambridge University Press, 2005.

\bibitem{Bai}
R.~W. Heath~Jr., M.~Kountouris, and T.~Bai, ``Modeling {H}eterogeneous
  {N}etwork {I}nterference {U}sing {P}oisson {P}oint {P}rocesses,''
  \emph{{IEEE} Trans. Signal Process.}, vol.~61, no.~16, pp. 4114--4126, August
  2013.

\bibitem{Marzetta1}
T.~Marzetta, ``Noncooperative {C}ellular {W}ireless with {U}nlimited {N}umbers
  of {B}ase {S}tation {A}ntennas,'' \emph{{IEEE} Trans. Wireless Commun.},
  vol.~9, no.~11, pp. 3590--3600, November 2010.

\bibitem{Haenggi}
M.~Haenggi, \emph{Stochastic {G}eometry for {W}ireless {N}etworks}.\hskip 1em
  plus 0.5em minus 0.4em\relax Cambridge University Press, 2013.

\end{thebibliography}


\end{document}